\documentclass[review]{elsarticle}

\usepackage{lineno,hyperref}
\modulolinenumbers[5]

\journal{Speech Communication}

\usepackage{newtxtext,newtxmath}
\usepackage{booktabs}

\begin{document}

\begin{frontmatter}

\title{Controllable speech synthesis by learning discrete phoneme-level prosodic representations}

\tnotetext[t1]{Preprint - the final authenticated publication is available online at: \url{https://doi.org/10.1016/j.specom.2022.11.006}}
\tnotetext[t2]{$\copyright$ 2022. This manuscript version is made available under the CC-BY-NC-ND 4.0 license \url{https://creativecommons.org/licenses/by-nc-nd/4.0}}


\author[segraddress,uniwaaddress]{Nikolaos Ellinas\corref{mycorrespondingauthor}}

\cortext[mycorrespondingauthor]{Corresponding author}
\ead{n.ellinas@samsung.com}

\author[segraddress]{Myrsini Christidou}
\author[segraddress]{Alexandra Vioni}
\author[hqaddress]{June Sig Sung}
\author[segraddress]{Aimilios Chalamandaris}
\author[segraddress]{Pirros Tsiakoulis}
\author[uniwaaddress]{Paris Mastorocostas}


\address[segraddress]{Innoetics, Samsung Electronics, Greece}
\address[hqaddress]{Mobile Communications Business, Samsung Electronics, Republic of Korea}
\address[uniwaaddress]{Dept. of Informatics \& Computer Engineering, University of West Attica, Greece}

\begin{abstract}
In this paper, we present a novel method for phoneme-level prosody control of F0 and duration using intuitive discrete labels.
We propose an unsupervised prosodic clustering process which is used to discretize phoneme-level F0 and duration features from a multispeaker speech dataset.
These features are fed as an input sequence of prosodic labels to a prosody encoder module which augments an autoregressive attention-based text-to-speech model.
We utilize various methods in order to improve prosodic control range and coverage, such as augmentation, F0 normalization, balanced clustering for duration and speaker-independent clustering.
The final model enables fine-grained phoneme-level prosody control for all speakers contained in the training set, while maintaining the speaker identity.
Instead of relying on reference utterances for inference, we introduce a prior prosody encoder which learns the style of each speaker and enables speech synthesis without the requirement of reference audio.
We also fine-tune the multispeaker model to unseen speakers with limited amounts of data, as a realistic application scenario and show that the prosody control capabilities are maintained, verifying that the speaker-independent prosodic clustering is effective.
Experimental results show that the model has high output speech quality and that the proposed method allows efficient prosody control within each speaker's range despite the variability that a multispeaker setting introduces.
\end{abstract}


\end{frontmatter}


\section{Introduction}
\label{sec:intro}

Since state-of-the-art neural text-to-speech (TTS) systems have achieved high speech quality and naturalness in both single speaker \cite{wang2017tacotron, shen2018natural} and multispeaker \cite{ping2018deep, zhang2019learning} setups, more specialized tasks concerning the prosody of synthesized speech have arisen as major challenges.
Some of these include prosody transfer, prosody control and even the production of expressive speech based on human emotions.
In this work, we focus on the task of prosody control/manipulation and specifically the representation of two important aspects of prosody, F0 and duration at the phoneme level.

\subsection{Related Work}

The first approaches towards manipulating the prosody of synthesized speech in a neural TTS system consist of global prosody embeddings \cite{Skerry-Ryan2018}.
These trainable embeddings are extracted from the utterances of the training set and aim at transferring the utterance-level prosody of a reference audio to the synthesized speech.
A milestone in this domain is the introduction of Global Style Tokens (GSTs) \cite{GSTs}, a codebook learned in an unsupervised manner, which allows control of many different aspects of prosody by using a weighted sum of disentangled style embeddings.
It is also shown that Variational Autoencoders (VAEs) can very effectively learn latent prosodic representations in an unsupervised way.
These representations can either be learned with a single Gaussian distribution \cite{Zhang2019} or even with a Gaussian mixture model by modeling it as two hierarchical levels of latent variables \cite{hsu2018hierarchical}.
Such a model can also effectively incorporate the capacity of the learned embeddings \cite{battenberg2019effective}, which measures the amount of information contained in the latent variables, or even be adapted to a semi-supervised setting \cite{habib2019semisupervised}.

Apart from the general style that the utterance-level approaches can offer, a more fine-grained approach can be also achieved in multiple resolutions such as word-level, phoneme-level or even frame-level.
In \cite{Lee2019} variable-length speech prosody embeddings are learned, allowing frame-level control of pitch and energy.
Text-side prosody embeddings are also learned by first aligning the reference spectrogram and the corresponding utterance phonemes with the help of an attention mechanism.
By utilizing the latter method, more sophisticated methods are possible such as representing word and phoneme-level aspects of prosody as a hierarchical fine-grained variable model \cite{Sun2020}.
\cite{Sun2020a} shows that discretizing phoneme-level latent features and using an autoregressive prior generates more natural samples instead of simply sampling from a standard VAE prior.
Extending the idea of GSTs, a pretrained wav2vec 2.0 \cite{baevski2020wav2vec} model can be used to capture local style patterns in a transformer-based architecture \cite{chen2021fine}.
In \cite{klapsas2021word}, prosody is controlled by incorporating a word-level GST module in a non-attentive Tacotron model \cite{shen2020non}, with the addition of an autoregressive prior which allows high quality speech synthesis without requiring a reference audio.

Explicit prosodic features such as phoneme duration, F0 or energy can be efficiently calculated by modules which are based on common signal processing methods.
These features can be used as labels in a supervised setting as they contain rich information that can facilitate convergence and provide direct control over the corresponding aspect of prosody.
Integrating phoneme durations extracted from a phoneme aligner in a Tacotron architecture enables phoneme-level duration control during inference \cite{Park2019}.
Non-attentive Tacotron \cite{shen2020non} and FastSpeech \cite{ren2019fastspeech} introduce non-autoregressive TTS architectures with integrated phoneme-level duration control.
Combining explicit duration information in a speaker embedding network can also improve the rhythm of synthesized utterances in a multispeaker setup \cite{fujita21_interspeech}.

Extending to other prosodic features, FastPitch \cite{lancucki2021fastpitch} incorporates pitch control by also predicting F0 contours and FastPitchFormant \cite{bak21_interspeech} utilizes the predicted F0 in an excitation generator inspired by the source-filter theory in order to provide more robust and accurate pitch control.
Since TTS decoders are conditioned on phoneme encoder representations, in FastSpeech 2 \cite{ren2021fastspeech} and FCL-Taco2 \cite{wang2021fcl} prosody prediction modules are introduced, which add prosodic information to these representations and are trained in a supervised manner utilizing ground truth values.
In these cases, prosody information can be represented in various ways.
For example, FastSpeech 2 uses the Continuous Wavelet Transformation for F0, whereas \cite{du2021mixture} uses Mixture Density Networks.
\cite{chien2020hierarchical} utilizes handcrafted prosodic features from open-source packages as well as neural features learned via vector quantization to extend FastSpeech 2 for improved quality and prosody naturalness.

Mellotron \cite{valle2020mellotron} combines GSTs and explicit F0 values to guide a Tacotron 2 decoder and achieves variable style synthesis as well as singing.
F0 and other real-valued prosodic features can also be aggregated per phoneme and enable fine-grained prosody transfer \cite{Klimkov2019} and control \cite{mohan21_interspeech}, whereas in \cite{gong2021improving} F0 is represented in a discrete way using a vector quantization module.

The phonetic encoder outputs are linguistic representations of the input text, so they can be used to predict the expected prosody of the synthesized utterance.
This can be achieved either by predicting prosodic features directly from the encoder outputs \cite{Shechtman2019,Raitio2020}, or preprocessing the features with a reference encoder and use its output representation \cite{Gururani2019}.
In \cite{Wan2019}, acoustic and linguistic representations are combined in multiple resolutions such as syllable, word or sentence level in order to predict F0, duration and energy as well as capture latent prosodic variations.
Other text-derived features, such as part-of-speech tags or BERT embeddings can also be used to adapt prosody according to linguistic context \cite{9414413,9413696}.

In multispeaker setups, fine-grained processing of prosodic features is shown to disentangle speaker and style information.
Hence, it can be successfully used in tasks such as voice conversion \cite{wang2021adversarially} and personalized TTS systems \cite{9414422,kumar2020shot}.
Leakage between speaker and style information must also be avoided in prosody transfer scenarios.
Fine-grained representations extracted with a variational reference encoder combined with speaker embeddings \cite{karlapati2020copycat}, as well as collaborative and adversarial learning \cite{daxin2020finegrained} are successful in disentangling speaker and style content.
Finally, apart from manipulating the aforementioned prosodic features, naturalness of synthesized speech can be improved by focusing on higher-level characteristics such as stress or intonation.
This can be achieved by adapting the BERT framework \cite{kenton2019bert} for speech synthesis \cite{chen2021speech} or utilizing ToBI linguistic prosodic labels \cite{kimtobi,zou21_interspeech}.

\subsection{Proposed Method}

In this paper we present a method for phoneme-level prosody control of F0 and duration.
Extensive research has been performed on conditioning TTS models on prosodic features, such as the continuous F0 contour at frame level \cite{valle2020mellotron}.
Other real valued features such as duration and energy can also be used to condition a Tacotron decoder.
The features used are aligned per phoneme and aggregated either by using an attention module \cite{Klimkov2019} or simple forced alignment \cite{mohan21_interspeech}.
Real valued features can impose some limitations due to their one-dimensional nature.
In FastSpeech 2 \cite{ren2021fastspeech}, the continuous wavelet transform is applied for representing the F0 as a spectrogram, whereas in \cite{gong2021improving} discrete F0 representations using vector quantization are shown to improve pitch-related prosody compared to their continuous counterpart.

In our work, we use discrete representations as they provide a form of regularization over the prosodic features and are shown to increase naturalness while maintaining appropriate diversity \cite{Sun2020a,gong2021improving}.
Instead of training a quantized fine-grained VAE in order to learn latent representations, we simply use extracted features such as F0 and duration, the values of which are determined by standard speech processing tools.
The discretization is then performed at the phoneme level using simple clustering methods, such as K-Means clustering, resulting in humanly interpretable labels which are directly applied to the dataset without requiring training \cite{Vioni2021}.
An additional group of encoder and attention modules learn to model the discrete sequences and disentangle their content from the corresponding phoneme-level linguistic features.

The learned labels provide great controllability in synthesized speech, however they are bounded by the speaker's range, since the outermost clusters may contain extreme values which are not frequent in the training data.
The proposed method is directly applied to multispeaker TTS and enables phoneme-level prosody control for every speaker included in the training set \cite{Christidou2021}.
We also introduce a prosody predictor module to enable end-to-end TTS without the need of reference audio or manually selected labels.
Using ordinal labels with a limited range, e.g. 1 up to 15, in a multispeaker setup is a direct and intuitive way of controlling the prosody, while also provides an advantage over continuous values as it does not require prior knowledge of acceptable real valued intervals for each speaker in the dataset.

Several preprocessing steps are applied in order to obtain effective and meaningful representations.
For F0 values, we apply per speaker normalization and K-Means clustering with the purpose of neutralizing speaker and gender variations and obtain speaker-independent labels which explain the same space.
A balanced duration clustering strategy is used for the durations of each phoneme separately, assigning an equal number of samples in each cluster in order to address instabilities due to cluster imbalance, a problem which was found in our previous work \cite{Vioni2021} and we would like to alleviate.
Augmentation transformations are also applied to the training data \cite{angelini2020singing}, in order to increase the number of samples in the outermost clusters.

The final end-to-end system allows us to control F0 and duration at the phoneme level by creating universal, speaker-independent clusters.
The same model is directly used for speaker adaptation \cite{markopoulos21_ssw}, extending the TTS and prosody control capabilities to previously unseen speakers after fine-tuning the model with only a few samples.
Introducing a prosody predictor module, which is trained separately and in minimal time after the main model has finished training, we also increase the flexibility of the model by enabling end-to-end TTS without requiring manually selected labels or a reference audio.
The prosody predictor output can be used as is for the default rendering of any utterance, or it can be manipulated according to the desired control specification at the phoneme or any higher level.

As an additional note, in this work we consider F0 and duration as the main factors of prosody in order to present the effectiveness of our method.
We leave energy and other factors as future work, where their role will be more crucial in the final result, e.g. singing synthesis.

We could summarize the contributions of our work with the following:
\begin{itemize}
	\item Prosodic clustering for fine-grained phoneme-level prosody control
	\item Controllable end-to-end text-to-speech synthesis using intuitive discrete labels
	\item Multispeaker prosody control, with application to unseen speaker adaptation
\end{itemize}

\section{Method} 
\label{method}

\subsection{Data preprocessing}

The acoustic model uses phonemes as linguistic inputs which are produced by a front-end preprocessing module from the input text.
An HMM monophone acoustic model is used as a forced-alignment system \cite{raptis2016expressive} to obtain accurate alignments between the utterance and its corresponding phonetic transcription.
The duration of each phoneme is extracted from the alignments, excluding word boundaries and pauses although they are included in the phonetic sequence to be modeled by the acoustic model.


A standard autocorrelation method is used \cite{boersma1993accurate} for F0 extraction followed by smoothing and interpolation in the unvoiced regions, as well as a transformation to the logarithmic domain.
The proposed model uses phoneme-level F0 features, so the previously extracted alignments were utilized to obtain each phoneme's corresponding log-F0 values.
Then, in order to account for the variable phoneme duration, these values are averaged.
From this point on, we will refer to as F0 features the mean log-F0 values of each phoneme.

Aiming to make our model more robust, voice data augmentation is also applied in order to widen the speaker range and to increase the number of samples in each cluster, as seen in previous multispeaker \cite{Cooper2020} and singing synthesis \cite{blaauw2017,angelini2020singing} papers.
The twelve data transformations applied are: pitch shifting by [-6, -4, -2, 2, 4, 6] semitones, and tempo changes by altering speaking rate to [0.70, 0.80, 0.90, 1.10, 1.20, 1.30] of the original one, using the Praat Vocal Toolkit \cite{corretge2012pvt}.
As mentioned in \cite{angelini2020singing}, a small augmentation does not significantly degrade speech quality in singing datasets, so we consider safe to also apply these transformations to our dataset since it contains plain speech which does not have extreme values in terms of pitch or speaking rate.
The differences will be evened out more when the quality degradation from the acoustic model and the vocoder are taken into account.
In our case, we do not want the augmentation to be excessive, but only adequate to make the clustering process more effective, so we did not overlap the transformations.
Hence, we applied the augmentations in our dataset uniformly by splitting the dataset randomly in twelve sets and applying one transformation on each set.
The resulting dataset was double in size compared to the original one.
We perform clustering on our augmented dataset together with the original one to get cluster centroids that correspond to the widened F0 and duration ranges.
The proposed augmentation method extends the prosodic range, while also enhancing model robustness and voice quality.

\subsection{Prosodic clustering}

\begin{figure}[t]
	\centering
	\includegraphics[width=.6\textwidth]{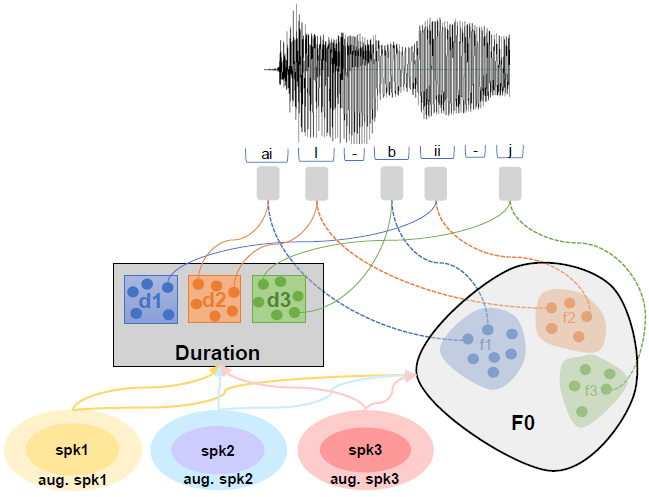}
	\caption{Clustering of prosodic features for multiple speakers.
		The normalization applied to each cluster is common for every speaker.}
	\label{fig:architectureclust}
\end{figure}

K-Means clustering with the squared distance criterion is applied for F0.
The resulting centroids can be translated as the representative values for F0 and can be used as a vocabulary of discrete tokens.
Since we are interested in a multispeaker setting, where different speakers have different pitch ranges, it is not practical to cluster the F0 values of each speaker separately.
In order to handle this, we first apply $z$-score normalization to each speaker's extracted F0 values, a common method widely used in similar speech research, such as ASR and multispeaker TTS.
We then cluster the normalized values of all speakers together to obtain universal F0 centroids. 
Thus for each phoneme's corresponding pitch in the dataset we apply: 

\begin{equation}
	F_0 =\frac{f-\mu_i}{\sigma_i} , i \in speakers
\end{equation}

\noindent where $f$ is the unnormalized F0 and $\mu$, $\sigma$ are the mean and variance of the respective speaker's F0 values.
This way, we deal with gender and speaker variation in pitch and create a mapping from each speaker's F0 values to a common prosodic space, where clustering can be performed universally.
The normalization method also facilitates adding new speakers, because the new F0 values can be directly mapped to the universal centroids without the need of recomputing them.

For the duration feature, clustering is performed separately per phoneme as phoneme classes, such as vowels and consonants, differ substantially depending on their articulation characteristics.
Results from our previous work \cite{Vioni2021} show that voice quality deterioration when using the outermost clusters is not so severe in F0 control compared to duration control.
Thus, we have adopted a balanced clustering method for extracting duration clusters.
The average phoneme duration values of all speakers are sorted in ascending order and grouped into the desired number of intervals, so that each interval contains an equal number of samples.
We observe that using this grouping strategy slightly decreases the duration control range, as extreme values are averaged out by being pushed towards the bulk of more frequent phoneme durations, but more importantly increases the duration control stability.
We note that our dataset contains speakers that have similar speaking rates and each speaker has a consistent speaking rate across the dataset, thus no duration normalization is necessary, as phoneme durations are similar.
It must be noted that the training dataset was not specifically designed to have speakers with similar speaking rates.
Moreover, the speakers that were selected for speaker adaptation were not selected based on the speaking rate, but rather include two randomly chosen internal speakers that were held out as well as benchmark open domain voices.
In other cases though, where the dataset is more diverse with more varying speaking rates, normalization might be necessary.

At training time, for each phoneme in an utterance, its corresponding prosodic feature is assigned to the nearest cluster centroid, resulting in a sequence of prosodic labels.
Each label is represented by an embedding vector, so that a sequence similar to the phoneme input sequence is produced, which can condition the decoder.
An overview of the method is presented in Figure~\ref{fig:architectureclust}.

\subsection{Acoustic model architecture} 
\label{multispeaker_model}

\begin{figure}[t]
	\centering
	\includegraphics[width=\textwidth]{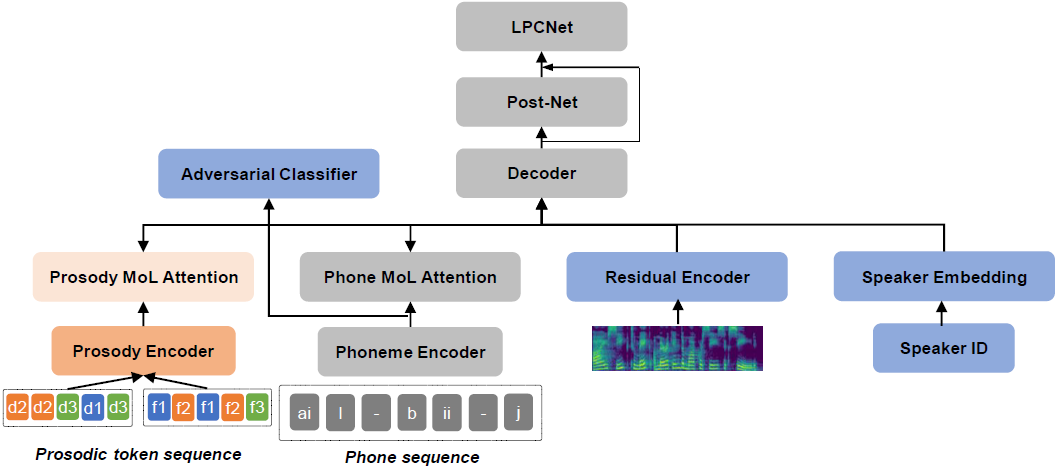}
	\caption{Multispeaker prosody control model.
		The single speaker architecture was modified with an Adversarial Classifier, Speaker Embedding and Residual Encoder in order to enable the model to train on multiple speakers.}
	\label{fig:architecture}
\end{figure}

The acoustic model is based on our previous work \cite{Ellinas2020,Vioni2021} adapted to a multispeaker architecture \cite{Christidou2021}.
It is an autoregressive attention-based text-to-speech model, that receives an input sequence of phonemes $\boldsymbol{p}=[p_1,...,p_N]$ and sequences of F0 and duration tokens $\boldsymbol{f}=[f_1,...,f_M]$, $\boldsymbol{d}=[d_1,...,d_M]$ which are jointly referred to as prosodic features.
Training on multiple speakers is enabled by including a speaker embedding layer, an adversarial speaker classifier and a residual encoder.
The full architecture is presented in Figure~\ref{fig:architecture}.

Each phoneme has a corresponding token for F0 and duration, while word boundaries and punctuation marks do not receive any such tokens, therefore ${M<N}$. 
The phoneme sequence is passed into a text encoder which produces a text encoder representation $\boldsymbol{e}=[e_1,...,e_N]$ and the prosodic feature sequences are concatenated and then passed into a prosody encoder which produces the prosody encoder representation $\boldsymbol{e'}=[e'_1,...,e'_M]$. 
On the decoder side, the attention RNN produces a hidden state $h_i$ which is used as a query in the attention mechanism for calculating the context vector $c_i$ representing phoneme information.
In our method, a prosodic attention context vector $c_i'$ is also produced and conditions the autoregressive decoder at each timestep, allowing the phoneme and prosody information to be modeled separately.
As mentioned in \cite{Vioni2021}, we do not wish the prosodic attention context vector to contain any phoneme information, so we choose to process the prosodic sequence with a separate attention module which consumes the query $h_i$ and prosody encoder representations $\boldsymbol{e'}$.
The 2 context vectors are concatenated and along with the attention RNN hidden state are then fed to a stack of 2 decoder RNNs.
A simpler approach in which the phoneme and prosody representations are directly concatenated showed worse results in terms of quality and content disentanglement.

For the task of alignment, both in the case of phonemes and prosodic features, the Mixture of Logistics (MoL) attention module is chosen as it is capable of producing a robust phoneme alignment \cite{Ellinas2020}.
MoL attention is a purely location-based method and a direct variation of GMM attention \cite{graves2013generating} using logistic distributions \cite{vasquez2019melnet}.
This module ensures the monotonicity of the alignments and produces stable results \cite{gmmattention,Ellinas2020}.
The Cumulative Distribution Function (\ref{cdf}) of the logistic distribution is used to compute the alignment probabilities for each decoder timestep $i$ over each encoder timestep $j$ (\ref{scores}).

\begin{equation}
F(x;\mu,s)=\frac{1}{1+e^{-\frac{(x-\mu)}{s}}}=\sigma\left(\frac{x-\mu}{s}\right)
\label{cdf}
\end{equation}
\begin{equation}
a_{ij} = \sum_{k=1}^{K}w_{ik}\left(F(j+0.5;\mu_{ik},s_{ik})-F(j-0.5;\mu_{ik},s_{ik})\right)
\label{scores}
\end{equation}

The parameters of the mixture are calculated in (\ref{muik_wik}).
\begin{equation}
\mu_{ik}=\mu_{i-1k}+e^{\hat{\mu}_{ik}}
\quad\mathrm{}\quad
s_{ik}=e^{\hat{s}_{ik}}
\quad\mathrm{}\quad
w_{ik}=softmax(\hat{w}_{ik})
\label{muik_wik}
\end{equation}

The parameters $\hat{\mu}_{ik}$, $\hat{s}_{ik}$, $\hat{w}_{ik}$ are predicted by 2 fully connected layers which are applied to the attention RNN hidden state $h_i$ as shown in (\ref{mlp}).
\begin{equation}
\left(\hat{\mu}_{ik},\hat{s}_{ik},\hat{w}_{ik}\right)=W_2\tanh(W_1(h_i))
\label{mlp}
\end{equation}

The context vector is calculated as the weighted sum of the encoder representations (\ref{context}).
\begin{equation}
c_i = \sum_{j=1}^{N}a_{ij}e_j
\label{context}
\end{equation}

The output acoustic frames are predicted by a feed-forward layer and when the decoding is complete, the prediction is finetuned by a 5-layer convolutional post-net identical to \cite{shen2018natural}. Finally, a feed-forward gate layer predicts the stop token that signals the end of speech generation.

Each speaker is mapped to a 64-dimensional learnable embedding, which is used to condition the decoder.
A variational residual encoder \cite{hsu2018hierarchical} is implemented to model any additional information included in the audio samples other than speaker identity, text and prosodic features, like acoustic conditions and noise.
An adversarial speaker classifier similar to \cite{zhang2019learning} is also added, to induce disentanglement of the phoneme representations and the speakers' identity.

\subsection{Prosody predictor}

During inference, the proposed model would normally require predefined prosodic labels, either manually selected or extracted from a reference audio.
In order to perform arbitrary synthesis without this strict requirement, we train a separate module which learns to predict the F0 and duration labels from the phoneme encoder outputs.
Our method is similar to \cite{stanton2018predicting}, but instead of predicting global style token weights we leverage the discrete nature of the representations used in our model to directly predict the phoneme-level prosodic labels.

\begin{figure}[t]
	\centering
	\includegraphics[width=.9\linewidth]{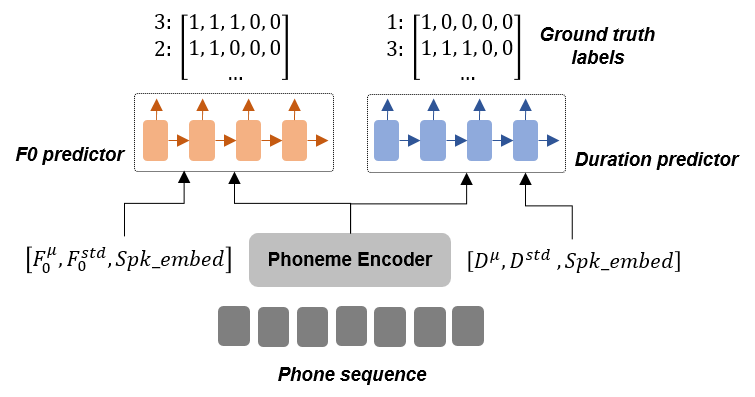}
	\caption{Prosody predictor module with ordinal regression targets. The weights of the encoder are kept frozen.}
	\label{fig:prospred}
\end{figure}

The prosody predictor module consists of two recurrent layers followed by a linear layer, mapping the output at each timestep to the corresponding categorical distribution of F0 and duration respectively, as also shown in Figure~\ref{fig:prospred}.
Since our proposed model is trained in a multispeaker setup, we also pass the speaker embedding as input to the prosody predictor.
Experiments also showed that including the global mean and standard deviation values of each F0 and duration sequence as additional input features helped achieve faster convergence and more stable predictions.
The training is performed using the cross-entropy criterion after the acoustic model is fully trained and while keeping its parameters frozen, so that the prosody prediction error does not backpropagate to the phoneme encoder.

Despite being discrete, the prosodic labels that we introduce have an ordinal structure, i.e. lower discrete values indicate a lower continuous value when translated back to F0 or duration.
In order to train the predictor module more efficiently, we adapt our learning procedure to account for ordinal categories.
This is implemented by assigning each label also to its lower-order categories, creating an incremental target vector of 'ones', instead of a one-hot vector for each category \cite{cheng2008neural}.
The \textit{sigmoid} function is also used for computing the probabilities instead of \textit{softmax}.

\subsection{Speaker adaptation}

The same method as described in the sections above was used to investigate the feasibility of fine-grained prosody control on a previously unseen speaker with a very small number of samples.
Extra attention was given so that the selected sentences, that were to be used in the new training, would provide enough phonetic coverage. 
This means that the utterances would contain each phoneme at least once. 

After applying augmentation and $z$-score normalization to the new speaker's data, we fine-tuned our pretrained model by replacing one of the speakers in the training set with the new speaker.
We experimented with various recording time lengths in order to test the model's limits and investigate how many minutes of recorded speech is needed to achieve similar quality results with the speakers in the training set.
We found that even with as few as 5 
minutes of recordings our model was able not only to reproduce that speaker's voice, but also to manipulate phoneme-level F0 and duration in a similar manner to the voices in our training set.

\section{Experiments}

In our experiments we aim to evaluate the speech quality and effectiveness of prosody control in a multispeaker model containing all available speakers and in a model adapted to a previously unseen speaker.
The evaluation of general prosody is a difficult task, as there are no consistent prosodic similarity or correctness metrics used in the current bibliography for arbitrary text-to-speech generation.
Thus, we focus on objective measurements and subjective evaluations of speech quality in order to verify our results.
We also assess the capability of the introduced prosody predictor module to generate prior information about the prosodic tokens, so that effective speech synthesis is possible without the need of predefined prosody labels.
In this case, we use the corresponding ground truth test utterances in order to measure prosodic correctness with commonly used metrics and directly compare with various cases, such as baseline, random labels and ground truth.

The basis for our experiments is a multispeaker internal dataset consisting of three female and two male voices, for a total of 159.7 hours of speech.
Additionally to this, the 2013 Blizzard Challenge Catherine Byers (Cathy) voice is obtained, in order to evaluate the effectiveness of proposed prosody prediction system in the developed speaker adaptation setup.
The preprocessing methods described in section \ref{method} are applied to all voice data, once for the internal multispeaker set alone and once with the addition of Cathy, leading to a total duration of 223.9 hours.
A randomly selected augmentation of either F0 or duration is applied to each utterance, creating new augmented speakers and doubling the size of the initial multispeaker dataset.
Grouping each speaker and their augmented version in a new set, $z$-score normalization of the extracted F0 values is applied for each one separately.
The duration labels are computed with balanced clustering, while the K-Means algorithm is used to find the optimal centroids for F0 values, with the selected number of clusters for both prosodic features being fixed to 15.
Since the duration values do not vary much amongst speakers and the F0 values are standardized, the values of each feature lie in the same space and the two clustering methods are applied on the whole augmented dataset with all the speakers mixed together.

\subsection{Speaker adaptation}

The internal multispeaker dataset together with Cathy is firstly used to train a model, referred to as \textit{Cathy-multi}, in order to generate samples from the target speaker when the speaker has been included in the initial training with the full set of its data.
For the adaptation model, \textit{Cathy-adapt}, we select 100 recordings from the Cathy dataset containing 7.72 minutes of speech and fine-tune the initial 5-voice multispeaker model.
In order to select a balanced group of recordings, the method introduced in \cite{Chalamandaris2009} is utilized to maximize the phonetic coverage in a small collection of recordings, by sorting the utterances of a speech corpus in descending order of phonological diversity.

To diversify the results in target speakers and genders, the initial multispeaker model is separately adapted to another 2 female and 2 male unseen voices by applying the speaker adaptation process independently for each one.
For this task, we use the LJ Speech dataset \cite{ljspeech17}, an audiobook male voice and two additional internal voices, one female and one male.
The corpus selection process to ensure phonetic coverage with 100 sentences for each voice resulted in 10.24, 5.7, 10.83, and 13.17 minutes of speech respectively.
By using these limited data we obtain the respective speaker adapted models, namely \textit{LJ-adapt}, \textit{Audiobook-male-adapt}, \textit{Female-adapt} and \textit{Male-adapt}.

The adaptation speakers pass through the same preprocessing steps for augmentation and prosodic clustering as described above, differentiating in the F0 centroids and duration intervals, which are not recomputed, but rather kept as obtained from the full length multispeaker dataset, in order to find the corresponding target speaker values.
The model is fine-tuned for 5K iterations as a single speaker model, after replacing one of the initial speaker identities with the target speaker to obtain the desired voice characteristics.

\subsection{Prosody prediction}

The prosody predictor module is trained after the multispeaker model has finished training and by freezing its weights.
The training utterances passed through the phoneme encoder, together with the corresponding speaker embeddings and ground truth prosodic labels form the training set of the prosody predictor.

For evaluating its performance we first train it on the \textit{Cathy-multi} model by using all of the training speakers, but for the evaluation we produce samples from the Cathy voice.
We also examine the speaker adaptation case, in which the predictor is firstly pretrained on the initial 5-speaker multispeaker model and then is fine-tuned on the \textit{Cathy-adapt} model using only the 100 adaptation utterances.
That way we ensure that the desired speaker remains unseen before the adaptation phase.

\subsection{Training Details}

All audio data was resampled at 24 kHz and the extracted acoustic features consist of 20 Bark-scale cepstral coefficients, the pitch period and pitch correlation, in order to match the modified LPCNet vocoder \cite{Vipperla2020}. The proposed model follows the same architecture as in \cite{Vioni2021} for the phoneme encoder, prosody encoder, attention mechanism and decoder with the additions described in \ref{multispeaker_model}.
The prosody predictor module consists of two 2-layered bidirectional LSTMs with 128 dimensions in each layer, followed by linear projections.
The Adam optimizer is used with weight decay with value $\lambda=10^{-6}$.
The predictor is trained in about 2 hours in the multispeaker case, and only 10 minutes in the adaptation case.

For the objective and subjective tests we selected 100 utterances from the dataset, which were excluded from the training data.
Those were used to extract the ground truth prosodic labels.

\section{Results}

\subsection{Objective evaluation}

In order to evaluate the control capability of the proposed model, a test set is generated for each speaker by assigning the prosody tokens of each sentence to a single cluster in an ascending order.
Specifically, this process is applied at one prosodic category at a time, keeping the other category's tokens at their ground truth values.

\begin{figure}[t]
	\centering
	\includegraphics[width=.9\linewidth]{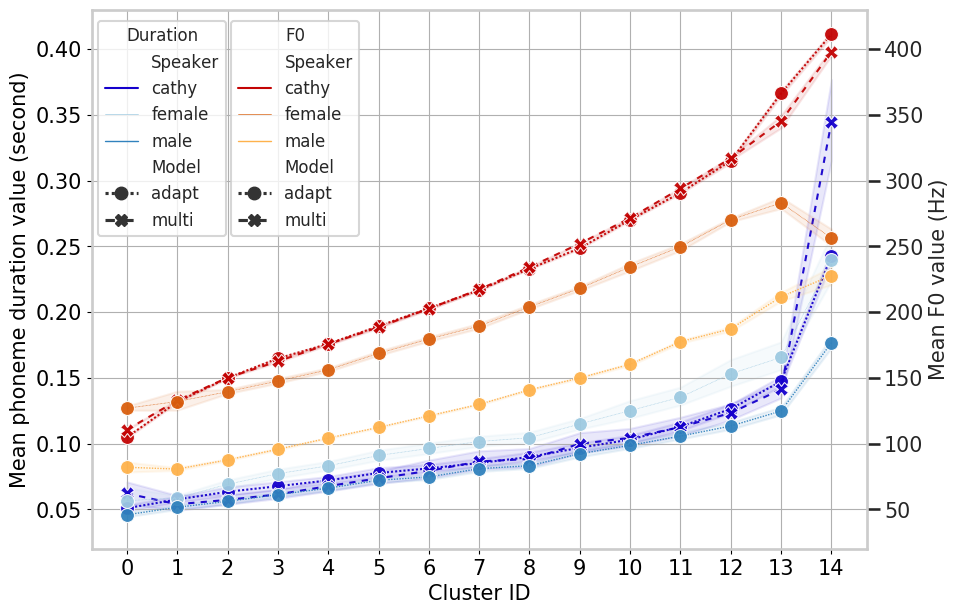}
	\caption{Sentence level mean F0 and phoneme duration for ascending cluster IDs with 95\% confidence intervals.
		The left y-axis corresponds to the duration graphs while the right y-axis corresponds to the F0 graphs.}
	\label{fig:ascending}
\end{figure}

In Figure~\ref{fig:ascending} the mean values of F0 and phoneme duration are depicted, calculated over the extracted features of every synthesized test utterance modified according to the specific cluster ID shown in the horizontal axis.
The depicted models belong to the configurations Cathy-multi, Cathy-adapt, Female-adapt and Male-adapt.
We can observe that all models follow the ascending order of the cluster IDs both in F0 and duration variations, proving that controllability is retained in the multispeaker setup.
Cathy-multi and Cathy-adapt perform alike and obtain similar values for the same cluster IDs, proving that prosody control in the same range is possible even with a few data, compared to a large speaker dataset.
The rest of the adaptation models present a same ascending behavior, with the male voices assuming lower F0 values from the female ones, despite being trained with a common set of prosodic labels, indicating that our method is indeed speaker and gender independent.

\begin{figure}[t]
	\centering
	\includegraphics[width=.8\linewidth]{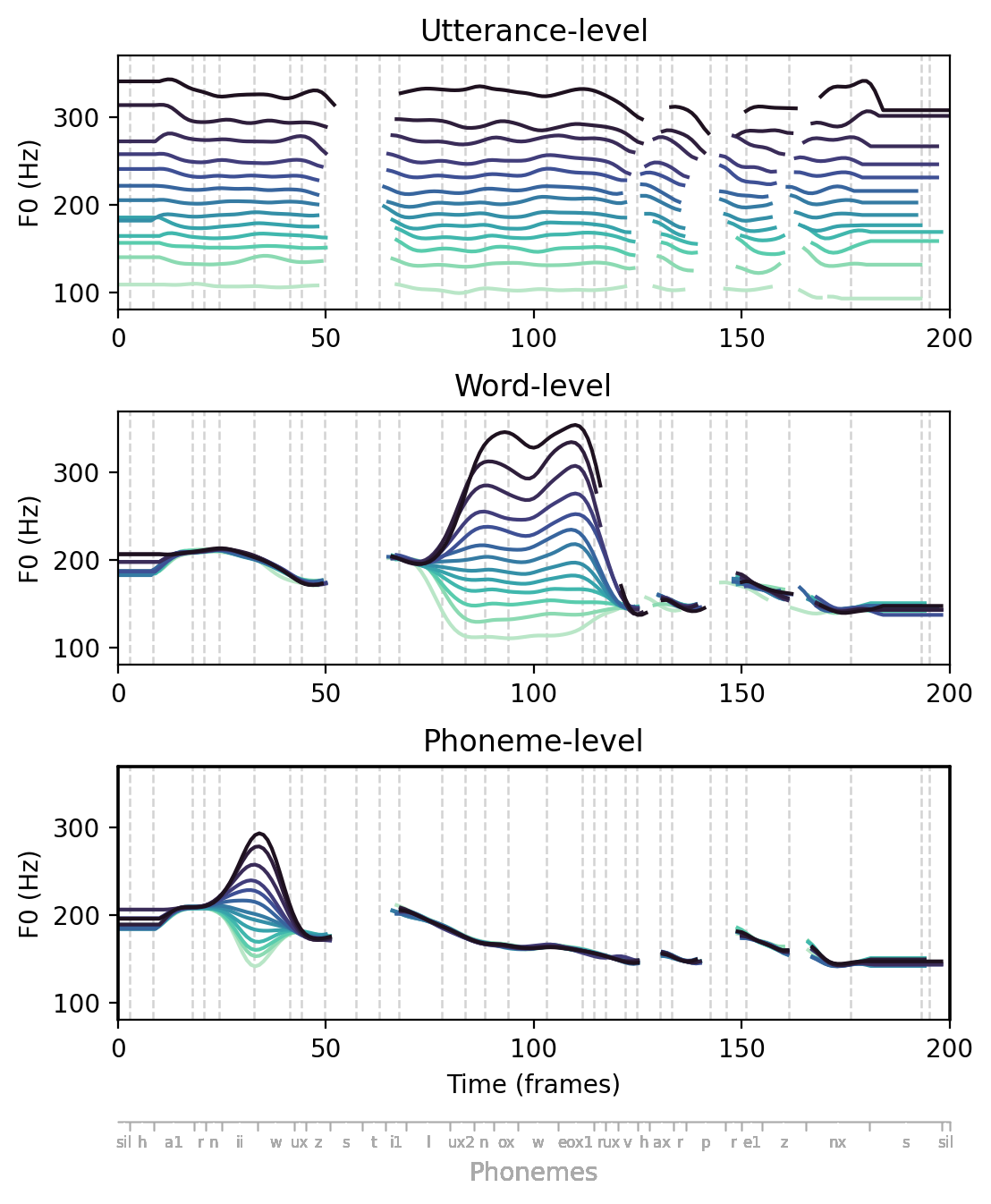}
	\caption{F0 contours of a test utterance, when varying the phoneme-level F0 control tokens for the whole utterance, a single word and a single phoneme.}
	\label{fig:utt-word-phoneme}
\end{figure}

Since we have shown the effectiveness of our method when varying the F0 and duration for the whole utterance, but no established protocol exists for word or phoneme-level evaluation, we encourage readers to listen to the audio samples on our web site\footnote{\scriptsize{https://innoetics.github.io/publications/prosodic-representations/index.html}}.
Also, in Figure~\ref{fig:utt-word-phoneme} a representative sample of the fine-grained control capability of our model over F0 is shown.
When varying the F0 tokens in the whole utterance, a single word or a single phoneme, the resulting F0 contours clearly depict the variations of F0 in the respective utterance segments.

\subsection{Subjective evaluation}

We performed listening tests in order to assess the quality of the proposed method, with respect to naturalness and speaker similarity.
Regarding naturalness, a set of 100 test sentences that were modified in terms of F0 and duration were used to synthesize voice samples from Cathy-multi, Cathy-adapt and Male-adapt models.
Listeners were asked to score the samples' naturalness on a 5-point Likert scale.

\begin{figure}[t]
	\centering
	\includegraphics[width=\linewidth]{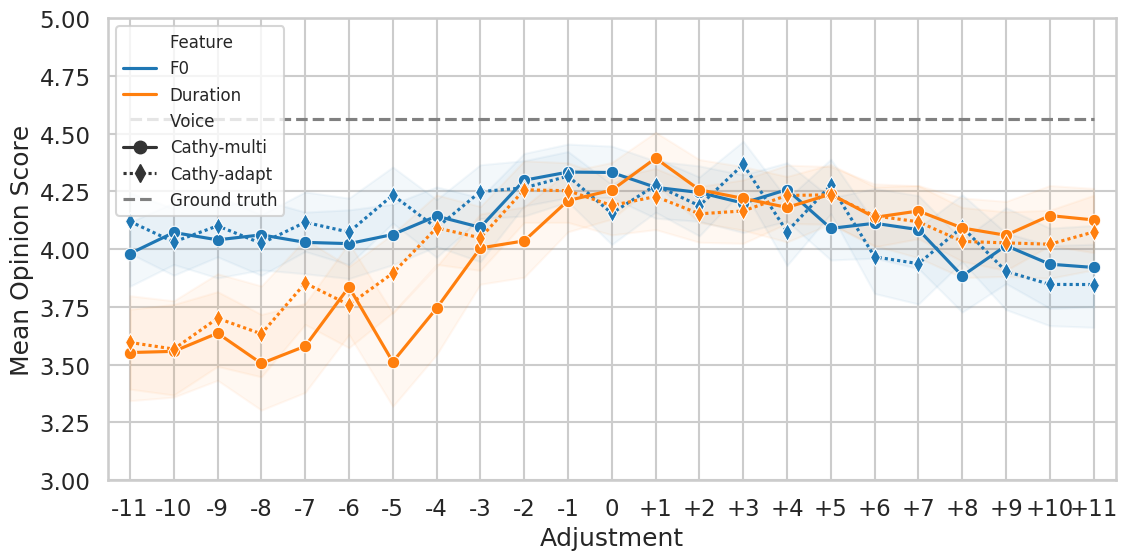}
	\caption{Mean opinion scores for Cathy-multi and Cathy-adapt with 95\% confidence intervals}
	\label{fig:mos-cathy}
\end{figure}

F0 and duration modification was done by adding or subtracting an offset from the ground truth prosodic labels of each test sentence, with the offsets varying in range $[-11,11]$.
Regarding F0, adding or subtracting an offset from the ground truth value leads to synthesized voice with higher or lower pitch, whilst regarding duration, these offsets lead to slower or faster uttered phones, respectively.
This method, which verifies the controllability of the model, is now used to evaluate naturalness.

In order to facilitate the Mean Opinion Score (MOS) results' visualization, each prosodic feature was modified independently, while the labels of the other feature retained their ground truth values.
In total, 2400 test utterances were rated for naturalness, with each one receiving 20 scores by native speakers via the Amazon Mechanical Turk.

\begin{figure}[t]
	\centering
	\includegraphics[width=\linewidth]{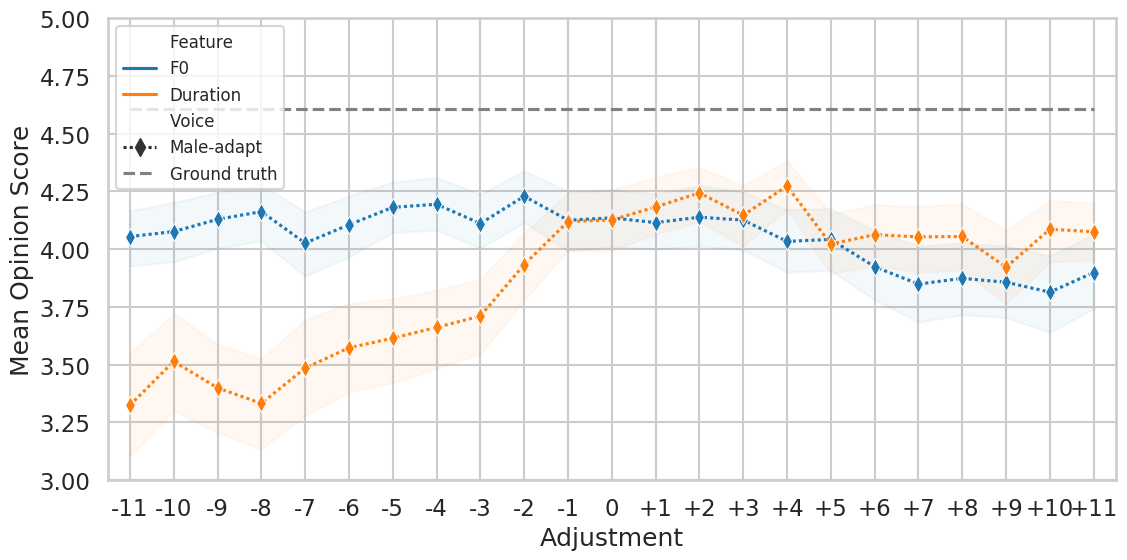}
	\caption{Mean opinion scores for Male-adapt with 95\% confidence intervals}
	\label{fig:mos-jsj}
\end{figure}

The MOS is depicted as a function of the modification offset in Figure~\ref{fig:mos-cathy} for the Cathy-multi and the Cathy-adapt models, and in Figure~\ref{fig:mos-jsj} for the Male-adapt model.
Based on the plots, it can be said that the voice samples with modified prosodic tokens retain reasonable naturalness levels in general, with the exception of very low duration offsets.
These offsets correspond to extremely fast speech which is generally considered unnatural.
Moreover, MOS scores of the voice samples produced by Cathy-adapt are directly comparable in naturalness with the scores of Cathy-multi, over the full modification range, as it can be seen in Figure~\ref{fig:mos-cathy}.
Hence, it is shown that, despite being trained with very limited data, the speaker-adapted models are capable of prosodic modification that also preserves high voice naturalness, in levels similar to the multispeaker model, which was trained with the full dataset.

Regarding speaker similarity, listening tests were performed to evaluate speaker adaptation with limited data.
For each speaker adapted model, 20 samples synthesized with ground truth prosodic labels were compared to a reference audio of the respective speaker.
Listeners were asked to rate speaker similarity on a 5-point Likert scale.
Each utterance received 40 scores by native speakers via the Amazon Mechanical Turk.

\begin{table}[b]
	\centering
	\begin{tabular}{l c}
		\toprule
		\textbf{Voice} & \textbf{Speaker Similarity} \\
		\midrule
		Cathy-adapt & $3.506 \pm 0.140$ \\
		LJ-adapt & $3.000 \pm 0.142$ \\
		Female-adapt & $3.858 \pm 0.124$\\
		Male-adapt & $3.942 \pm 0.117$ \\
		Audiobook-male-adapt & $3.633 \pm 0.121$ \\
		\bottomrule
	\end{tabular}
	\caption{Speaker similarity MOS for speaker adaptation with 95\% confidence intervals}
	\label{tab:spksim}
\end{table}

By observing the results in Table~\ref{tab:spksim}, it is evident that speaker similarity is adequate for all voices, taking into consideration that speaker adaptation was performed with only few minutes of speech from each speaker, and very satisfactory for the internal voices.

It can be said that the speaker similarity MOS scores correlate well with the voice recordings quality, since they are higher for internal voices with clear recordings, but deteriorate for voice datasets where noise and artifacts are present.

\subsection{Prosody predictor evaluation}

\begin{table}[b]
	\centering
	\begin{tabular}{@{\hspace*{1mm}} l @{\hspace*{3mm}} c @{\hspace*{3mm}} c @{\hspace*{3mm}} c @{\hspace*{3mm}} c @{\hspace*{3mm}} c}
		\cmidrule(ll){2-6}
		& \textbf{MOS} & \textbf{MCD} & \textbf{FFE} & \textbf{VDE} & \textbf{GPE} \\
		\midrule
		Predictor       & $4.380\pm0.060$                    & 5.8                     & 27.4                    & 6.9                     & 29.9                    \\
		Predictor-adapt & $4.396\pm0.059$                    & 6.2                     & 29.9                    & 7.1                     & 33.6                    \\
		GT labels       & $4.448\pm0.056$                    & 5.1                     & 8.2                     & 5.9                     & 3.5                     \\
		Random labels   & $3.858\pm0.083$                    & 6.6                     & 44.7                    & 9.5                     & 54.7                    \\
		Plain           & $4.364\pm0.060$                    & 5.8                     & 32.6                    & 6.7                     & 36.9                    \\
		\midrule
		Ground Truth    & $4.500\pm0.109$                     &                         &                         &                         &                      \\
		\bottomrule
	\end{tabular}
	\caption{Evaluation of prosody predictor module. For the objective metrics lower values are better.}
	\label{tab:prospredeval}
\end{table}

For evaluating the prosody predictor module we conducted listening tests on the selected 100 test utterances.
Listeners were asked to score the naturalness of 5 models.
We included the multispeaker prosody control model with the ground truth labels as the upper performance bound and the same model with random labels both for F0 and duration as the lower bound.
Also, a plain non-attentive model based on \cite{shen2020non} is included as a baseline, as it does not contain any prosodic information, but is considered one of the state-of-the-art systems for end-to-end TTS.
We also conducted an objective evaluation with commonly used metrics such as Mel-Cepstral Distortion (MCD) \cite{kubichek1993mel}, $F_0$ Frame Error (FFE), Voicing Decision Error (VDE) and Gross Pitch Error (GPE) \cite{chu2009reducing}.
These were calculated after the mel-spectrograms of the synthesized and ground truth sequences were aligned with dynamic time warping for each case separately.

Considering the results of Table~\ref{tab:prospredeval}, when using the ground truth labels the quality is the highest possible and very close to the ground truth, indicating that our proposed model produces high quality synthetic speech.
The quality of a plain model without any prosody control capabilities is shown to be lower.
This can be justified by the fact that our test voice is quite expressive, as it contains readings from books in an expressive style, hence the plain model cannot model all the prosodic variations and converges in a neutral overall style which may not match the corresponding ground truth utterances.
Regarding the prosody predictor results, we notice very high MOS scores both in multispeaker and speaker adaptation cases and lower FFE and GPE scores compared to the plain model.
This indicates that the F0 track is more similar to the ground truth in the case of the predictor models, confirming our previous conclusion about the plain model.
Finally, when using random labels the quality drops significantly in all measured aspects, proving that the predictor has learned the distributions of F0 and duration and that it is sufficiently trained.

\section{Conclusions}

In this paper, we presented an end-to-end controllable speech synthesis system that utilizes discrete phoneme-level prosodic representations based on F0 and duration clustering.
An additional encoder for the discrete prosodic representations along with a corresponding attention module were included in order to create alignments between the prosody encodings and the decoder hidden state.
We apply augmentation, feature normalization and a universal clustering method for all speakers' recordings so that we can produce universal F0 and duration representations for training.
The same principles are applied to new, previously unseen speakers with very few recordings, in order to test if this method can be used to create synthetic speech similar to the target voice with the same quality and level of control.

Our experiments verify that the multispeaker and speaker adapted models retain the control capability over F0 and duration and generate high quality speech, independently of gender or different voice characteristics.
Moreover, the speaker adapted models' scores indicate reasonable similarity to the original speakers' audio, given the short duration and variable quality of the recordings across speakers.

A prosody prediction module is also incorporated that predicts the discrete F0 and duration labels from the phoneme encoder outputs, enabling the model to produce synthetic speech which is close to the natural voice of each speaker even if specific labels are not given.
This module is shown to also be effective in the speaker adaptation scenario where it is fine-tuned for very few iterations on an unseen speaker, but produces meaningful prosodic labels.


\section*{References}

\bibliography{mybibfile}

\end{document}